\documentclass{article}

\usepackage[dblblindworkshop,final]{neurips_2026}
\workshoptitle{Beyond Private Training: The New Landscape of AI Privacy (InfPriv)}

\usepackage[utf8]{inputenc}
\usepackage[T1]{fontenc}
\usepackage{hyperref}
\usepackage{url}
\usepackage{booktabs}
\usepackage{amsfonts}
\usepackage{amsmath}
\usepackage{amssymb}
\usepackage{nicefrac}
\usepackage{microtype}
\usepackage{xcolor}
\usepackage[final]{graphicx}
\setkeys{Gin}{draft=false}
\usepackage{tikz}
\usetikzlibrary{arrows.meta,positioning,fit,backgrounds,calc}

\graphicspath{{./figs/}{figs/}}

\title{Beyond Output Filtering: \\Externally Auditable Erasure for\\Inference-Time RAG Memory}

\author{%
  Sean Culatana \\
  Atlassian \\
  Mountain View, CA \\
  \texttt{sculatana@atlassian.com} \\
  \and
  \textbf{Kang Li} \\
  Atlassian \\
  Bellevue, WA \\
  \texttt{kli3@atlassian.com}
}

\begin{document}
\maketitle

\begin{abstract} Retrieval-augmented systems increasingly rely on vector indexes that may retain deleted items in their search graph. Existing deletion interfaces can prevent deleted identifiers from appearing in returned results while still computing distances to their embeddings during graph traversal. We formalize this distinction as \emph{output safety} versus \emph{traversal safety}, and introduce \textsc{TSD-Audit}, a framework for auditing and enforcing traversal-safe deletion in graph-based approximate nearest-neighbor retrieval. On Faiss \texttt{IndexHNSWFlat}, native filtering leaves the number of distance computations unchanged relative to unfiltered search; at a $70\%$ deletion rate, trace-faithful replay detects deleted-vector scoring in all 100 audited queries. Code inspection of \texttt{hnswlib}'s \texttt{mark\_deleted} path reveals the same scoring-before-liveness pattern. \textsc{TSD-Audit} enforces an \emph{alive-before-scoring} invariant, repairs connectivity using only live candidates, and emits per-query scored-trace certificates that an independent verifier can check against the deletion snapshot. Under region-targeted deletion, \textsc{TSD-Audit} improves Recall@10 over native filtering by $4.3$--$42.2$ percentage points across deletion fractions from $0.5$ to $0.9$, while remaining comparable under random deletion. These results show that output-only deletion audits can miss process-level exposure: auditing deletion in vector retrieval requires accounting for the vectors scored during search, not only the identifiers returned. \end{abstract}
\section{Introduction}
Privacy research has historically anchored on the training phase, relying predominantly on differentially private training such as DP-SGD. But state-of-the-art applications increasingly pivot to inference-time, non-finetuning paradigms: autonomous agents, in-context learning, and retrieval-augmented generation. In these deployments a large share of the sensitive data an application handles never enters model weights at all---it sits in a \emph{retrieval memory} (a vector index over user documents, conversations, or tool traces) that is read at inference time. This relocates a core privacy obligation from training to deployment: when a user exercises a right to erasure, or an operator must expunge disallowed or poisoned content, the memory must delete the entry \emph{and} the system must be able to \emph{prove} the deletion.

\begin{figure}[t]
\centering
\begin{tikzpicture}[
  >={Stealth[length=2mm]}, line width=0.6pt,
  live/.style={circle,draw=black!65,fill=blue!8,minimum size=5mm,inner sep=0,font=\scriptsize},
  del/.style={circle,draw=red!75,fill=red!12,minimum size=5mm,inner sep=0,font=\scriptsize},
  ghost/.style={circle,draw=black!25,fill=black!4,minimum size=5mm,inner sep=0,font=\scriptsize,text=black!35},
  q/.style={circle,draw=black,fill=black!82,text=white,minimum size=5mm,inner sep=0,font=\scriptsize\bfseries},
  gedge/.style={-,black!45},
  scored/.style={->,red!78,dashed,line width=1pt},
  safe/.style={->,blue!62,line width=1pt},
  repair/.style={-,blue!55,dotted,line width=1pt},
]
\begin{scope}[local bounding box=L]
  \node[q]    (q)  at (0,0)      {q};
  \node[live] (a)  at (1.15,0.75){};
  \node[del]  (d)  at (2.35,0.35){d};
  \node[live] (b)  at (1.1,-0.85){};
  \node[live] (c)  at (3.15,-0.5){};
  \draw[gedge] (q)--(a); \draw[gedge](a)--(d); \draw[gedge](d)--(c);
  \draw[gedge] (q)--(b); \draw[gedge](b)--(c);
  \draw[safe]  (q) to[bend left=8] (a);
  \draw[scored](a) to[bend left=8] (d);
  \node[red!80,font=\scriptsize] at (2.35,-0.35) {scored};
\end{scope}
\node[font=\bfseries\small] at ($(L.north)+(0,0.28)$) {Soft-delete (output filter)};
\node[font=\scriptsize,align=center,text=red!80] at ($(L.south)+(0,-0.42)$)
  {returned leak $=0$ \textbf{but} visited leak $>0$\\[-1pt]\textbf{audit FAIL: deleted item was scored}};

\begin{scope}[xshift=6.4cm,local bounding box=R]
  \node[q]     (q2)  at (0,0)      {q};
  \node[live]  (a2)  at (1.15,0.75){};
  \node[ghost] (d2)  at (2.35,0.35){d};
  \node[live]  (b2)  at (1.1,-0.85){};
  \node[live]  (c2)  at (3.15,-0.5){};
  \draw[gedge] (q2)--(a2); \draw[gedge](q2)--(b2); \draw[gedge](b2)--(c2);
  \draw[repair] (a2) to[bend left=20] (c2);
  \draw[safe]  (q2) to[bend left=8] (a2);
  \draw[safe]  (a2) to[bend left=16] (c2);
  \node[black!45,font=\scriptsize] at (2.35,-0.35) {skipped};
  \node[blue!60,font=\scriptsize] at (2.15,1.02) {repair edge};
\end{scope}
\node[font=\bfseries\small] at ($(R.north)+(0,0.28)$) {Traversal-safe (alive-before-scoring)};
\node[font=\scriptsize,align=center,text=blue!62] at ($(R.south)+(0,-0.42)$)
  {returned leak $=0$ \textbf{and} visited leak $=0$\\[-1pt]\textbf{audit PASS: deleted item never scored}};
\end{tikzpicture}
\caption{\textbf{Returned vs.\ visited leakage.} A deleted memory entry $d$ is filtered from the returned top-$k$ under both regimes, so an output-level audit sees zero leakage in each. \emph{Left:} soft-delete still \emph{visits and scores} $d$ (dashed red), so the retrieval process reads erased data on nearly every query---invisible to an output audit. \emph{Right:} the \texttt{alive}-before-scoring invariant never admits $d$ to the candidate pool and an offline repair edge restores connectivity, driving visited leakage to zero and yielding an externally checkable certificate.}
\label{fig:concept}
\end{figure}
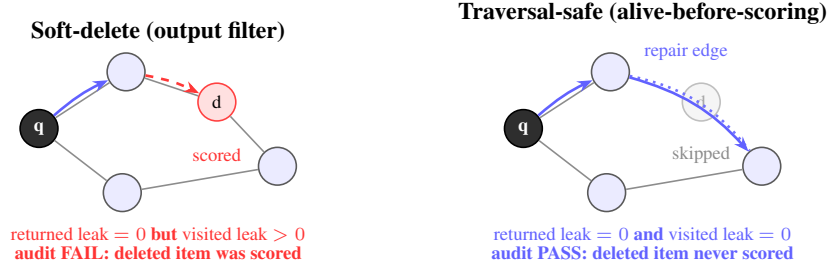

We study this obligation for the component most inference-time memories are built on: an approximate-nearest-neighbor (ANN) vector index. Production vector stores implement deletion as \emph{soft-delete}: a tombstone filters deleted ids from the returned top-$k$, but the graph search still visits and scores the deleted embeddings (Figure~\ref{fig:concept}). We argue this makes the standard, output-level compliance audit \emph{blind by construction}: it observes returned results, where leakage is zero, and never observes the retrieval process, where a deleted embedding is read on nearly every query. An assistant that ``forgot'' a user's revoked record at the output while still reading it internally on every query has not really forgotten it---and no output-level audit can tell the difference. We therefore elevate deletion from a returned-result constraint to an \emph{auditable retrieval-time guarantee}, and---unlike prior soft-delete analyses that use a reimplementation---establish the leak on the \emph{unmodified} \texttt{faiss} engine and emit a certificate an external party can check.

Our contribution is deliberately scoped as \emph{audit-not-algorithm}. The enforcement mechanism (an \texttt{alive} predicate checked before scoring) is the label-filtered traversal of predicate-aware ANN \citep{patel2024acorn,gollapudi2023filtereddiskann}; the recall-collapse-and-repair phenomenon is known churn behavior \citep{singh2021freshdiskann,xiao2024hnswrealtime}; and at-rest recoverability of soft-deleted vectors is studied by Ghost Vectors \citep{chakraborttii2026ghost}. What none of these own is a certifiable, externally verifiable guarantee that the retrieval \emph{process} never scored a deleted item---an emerging, inference-time privacy problem. We contribute: (1) the output-safe vs.\ traversal-safe distinction as measurable leakage; (2) the \texttt{alive}-before-scoring invariant and a traversal-safe repair; (3) an externally verifiable certificate of erasure; and (4) an \emph{external-validity audit on unmodified \texttt{faiss}} plus an \emph{honest baseline reconciliation} against \texttt{faiss}'s real soft-delete.

\section{Threat model and leakage}
We assume an \emph{honest-but-buggy} operator running an evolving inference-time pipeline (ANN retrieval feeding a reranker/LLM, with caches and logs). ``Compliant'' means no query-time operation reads or computes on a deleted item; ``auditable'' means this is checkable from emitted evidence without trusting the code path. We do \emph{not} target a malicious operator forging certificates, nor parameter-level unlearning of model weights \citep{bourtoule2021unlearning}. Let $\mathrm{Visited}(q,t)$ be every node on which the search computes a query--node distance. \textbf{Returned leakage} is the rate at which a deleted id appears in the top-$k$; \textbf{visited leakage} is the rate at which any deleted node is scored. A system is \textbf{traversal-safe} if both are zero. We stress the scope of the exposure: it is observable to a \emph{process-level} party (an auditor, or the operator's own telemetry). A purely client-observable adversary restricted to returned ids/scores is at chance---returned leakage is $0$---so the exposure is a process/compliance property of erasure, not a client-exploitable side channel. This is precisely the regime a deployment-time privacy audit must cover and an output test cannot.

\section{Method}
\textbf{Alive-before-scoring invariant.} The memory maintains an \texttt{alive} bitset consulted at three points, all \emph{before} a distance is computed: entry points are drawn from live nodes; deleted neighbors are dropped from the frontier before scoring; and a pop-time re-check skips nodes deleted after enqueue. Because deleted nodes never enter the candidate pool, $\mathrm{Visited}(q,t)\cap D_t=\emptyset$ holds for every query by construction. \textbf{Traversal-safe repair.} Pruning dead edges does not restore the connectivity deleted nodes provided; at high deletion the live subgraph fragments. An offline repair reconnects live nodes through paths that ran through deleted ones, scoring distances only among \emph{live} candidates, so it preserves the guarantee. \textbf{Certificate.} At each checkpoint we emit the deleted-set hash, hashes of sampled expansion traces, and the visited$\cap D_t$ verdict; a released verifier recomputes the hashes and asserts no-intersection, turning traversal-safe deletion into an externally checkable artifact.

\section{External validity on the unmodified \texttt{faiss} engine}
\label{sec:native}
To rule out that the leak is an artifact of our own beam search, we audit the \emph{native} \texttt{faiss}~1.11.0 HNSW search path (\texttt{IndexHNSWFlat}, $M{=}16$, $\mathrm{efC}{=}200$, $\mathrm{efS}{=}128$) over $114{,}516$ \texttt{bge} vectors, using \texttt{faiss}'s own \texttt{hnsw\_stats.ndis} counter---no custom code in the loop---at $70\%$ deletion ($3$ seeds), across three deployment conditions: (A)~\emph{soft-delete} (plain search, filter output); (B)~\emph{native selector} (\texttt{SearchParametersHNSW} with \texttt{IDSelectorBatch(alive)}); (C)~\emph{clean rebuild} on the live set.

The engine's own counter shows the built-in selector is output-safe, not traversal-safe: $\mathrm{ndis}(A){=}\mathrm{ndis}(B){=}1{,}535{,}152$ \emph{exactly}, every seed and both deletion modes (ratio $1.000$). \texttt{faiss}'s \texttt{IDSelectorBatch} does not prune before the distance kernel; deleted embeddings are still scored, only removed from output. Only rebuilding (C) stops the engine scoring them (ndis $1.05\times$ random, $1.38\times$ targeted). Returned leakage is $0$ in all conditions. We then emit the certificate from a replay whose fidelity to the native engine we validate directly (top-$10$ agreement $1.000$; replayed distance count $0.98\times$ \texttt{faiss}'s own \texttt{ndis}) and run the \emph{released} verifier over $100$ sampled queries: the soft-delete certificate \textbf{fails} (a deleted id in the scored set of every sampled query), while traversal-safe and clean-rebuild \textbf{pass} ($0$ hash/count mismatches; Table~\ref{tab:native}).

\begin{table}[h]
\centering\small
\caption{External-validity audit on unmodified \texttt{faiss} (70\% deletion, 3 seeds). The native selector leaves distance work unchanged; the released verifier fails soft-delete and passes the traversal-safe and rebuild certificates. Take-away: output filtering does not stop the memory from reading deleted entries.}
\label{tab:native}
\begin{tabular}{lccc}
\toprule
Condition & \texttt{ndis} (targeted) & returned leak & verifier \\
\midrule
A. soft-delete (output filter)      & $1{,}535{,}152$ & $0$ & \textbf{FAIL} \\
B. native \texttt{IDSelectorBatch}  & $1{,}535{,}152$ & $0$ & \textbf{FAIL} \\
C. clean rebuild (live only)        & $2{,}124{,}533$ & $0$ & PASS \\
Traversal-safe (ours)               & --- & $0$ & PASS \\
\bottomrule
\end{tabular}
\end{table}

\section{Honest recall reconciliation}
\label{sec:recall}
\begin{figure}[t]
\centering
\includegraphics[width=0.62\linewidth]{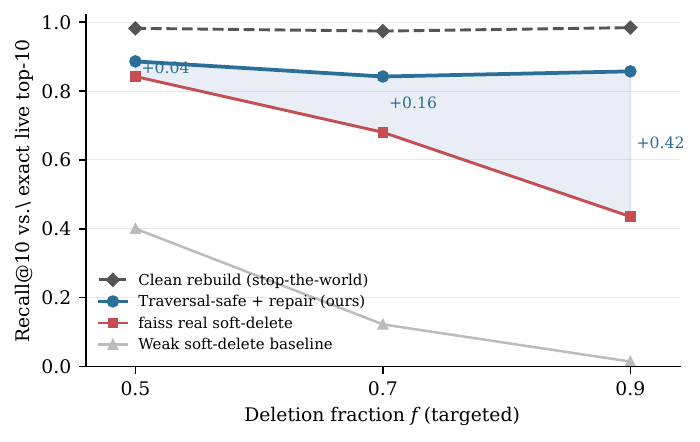}
\caption{\textbf{Recall@10 vs.\ exact live top-10 under targeted deletion} (mean of 3 seeds), against \texttt{faiss}'s \emph{real} soft-delete. The shaded band is the honest gap (traversal-safe$+$repair $-$ real soft-delete): $+0.043$/$+0.162$/$+0.422$ at $f{=}0.5$/$0.7$/$0.9$. Auditable erasure need not cost retrieval quality under targeted deletion; a stop-the-world rebuild leads on raw recall at rebuild cost.}
\label{fig:recall}
\end{figure}

A common but \emph{weak} soft-delete baseline oversamples a fixed $k'{>}k$ then filters; under targeted deletion this leaves few live results and collapses to near-zero recall. \texttt{faiss}'s real soft-delete (native selector, fill-$k$) is far stronger. We therefore reconcile our recall claims against \texttt{faiss}'s real soft-delete on the same graph (Figure~\ref{fig:recall}). Two honest findings. \textbf{(i)} Under \emph{random} deletion there is no quality advantage (all methods ${\approx}0.99$). \textbf{(ii)} Under \emph{targeted/region} deletion---the realistic case for erasing a user, a topic, or a compromised source---traversal-safe$+$repair beats \texttt{faiss}'s real soft-delete by a margin that \emph{grows} with the deletion fraction ($+4.3$/$+16.2$/$+42.2$ points at $f{=}0.5$/$0.7$/$0.9$), far smaller than the gap against the weak baseline. A stop-the-world clean rebuild attains the highest raw recall ($\geq 0.97$) at every fraction; traversal-safe$+$repair is therefore best read as an \emph{online, availability-preserving, certificate-clean} recovery that beats real soft-delete, not as a replacement for a rebuild on raw quality.

\section{Related work and scope}
Graph ANN \citep{malkov2016hnsw,subramanya2019diskann} in libraries such as \texttt{faiss} \citep{douze2024faiss} and dynamic benchmarks \citep{simhadri2025bigann} make dynamic retrieval practically important; recent work studies deletion algorithmically \citep{xu2025ipdiskann} and as evaluation \citep{yamashita2025deletioneval}. We concede the alive-filter mechanism to predicate-filtered ANN \citep{patel2024acorn,gollapudi2023filtereddiskann}, the recall-collapse-and-repair phenomenon to the churn literature \citep{singh2021freshdiskann,xiao2024hnswrealtime}, and at-rest recoverability to Ghost Vectors \citep{chakraborttii2026ghost}. Verifiable-ANN certifies result \emph{correctness} \citep{wang2025verifiableann} and factual-residual audits operate at the model level \citep{raeesi2026forgetaudit}; neither certifies the retrieval process of an inference-time memory. Our residual, defensible contribution is the traversal-exposure \emph{audit property} for erasure and its externally verifiable certificate.

\section{Limitations}
The exposure is a process/compliance property, not a client-exploitable channel (returned leakage is $0$). Our quality benefit is scoped to targeted/region deletion and grows with $f$; under random deletion it vanishes, and a stop-the-world rebuild dominates raw recall at rebuild cost. Certificates are scoped to an honest-but-buggy operator; a malicious operator forging traces requires a trusted execution environment or transparency log, which we leave to future work. We study the RAG memory layer specifically; auditable erasure for parametric or in-context memory is complementary and out of scope.

\end{document}